\documentclass[twocolumn,preprintnumbers,amsmath,amssymb]{revtex4-1}

\def\be{\begin{equation}}
\def\ee{\end{equation}}
\def\bea{\begin{eqnarray}}
\def\eea{\end{eqnarray}}
\def\e{\epsilon}

\def\m{\mu}
\def\n{\nu}

\def\p{\partial}
\def\a{\alpha}
\def\b{\beta}
\def\t{\theta}

\def\i{\imath}

\usepackage{dcolumn}
\usepackage{epsfig}
\usepackage{amssymb,amsmath,bm,graphicx,multirow,mathtools}
\usepackage{color}
\usepackage{epsfig}
\usepackage{bm}

\begin{document}

\title{The exact solutions to two-dimensional Yang-Mills theory: $su(2)$ and $u_\star(1)$}

\author{Abolfazl\ Jafari\footnote[1]{jafari-ab@sci.sku.ac.ir}}
\affiliation{Department of Physics, Faculty of Science,
Shahrekord University, P. O. Box 115, Shahrekord, Iran}
\date{\today }

\begin{abstract}
\begin{center}
\center{\textbf{Abstract}}
\end{center} 
The purpose of this study is to show that the exact solutions to Yang-Mills theory can occur in two-dimensional space-time. 
We show that the instability of the stationary solutions for the nonabelian gauge field theories 
questioned by the dimensions of space-time.
We show that instability of the stationary solution for the nonabelian gauge field theory questioned by the dimension of space-time.
Another intention of this work is research of the source-free solutions to Maxwell's equations in a two-dimensional noncommutative temporal state space-time.
We will show that there is comprehensively similarity between the existence of the bound states of nonabelian gauge theory and $u_\star(1)$.
\end{abstract}


\maketitle


\section{Introduction}

Derrick's theorem is an argument proposed by a physicist G.H. Derrick, who has shown that the
instability of the stationary localized solutions for a nonlinear
wave equation or the nonlinear Klein-Gordon equation in dimensions three and higher \cite{derrick,kolokolov, raja, dod}.
Also, Coleman and Deser have published the cardinal theorem about
the absence of time independent solutions for the source-free Yang-Mills equation:
"There are no static solutions for the source-free Yang-Mills field's equation of motion" \cite{coleman, deser}.
Instability of the static solutions, in particular, is satisfactory in quantum chromodynamics.
This idea can be generalized to include quantum electrodynamics entirely.
Perhaps, one of the consequences of the lemma is the absence of classical glueballs.
In nonabelian gauge theory, there are some kinds of bound states which are so-called the glueballs of QCD \cite{hou-wong}.
The glueballs as colorless bound states of the massive composed
exclusively of the gauge field quanta (gluons) which gets the mass in the Higgs mechanism.
In Refs.\cite{coleman, deser}, Deser and Coleman have shown that the classical glueballs cannot occur in the space-time with $1+1$ and $3+1$-dimensions.
A Lump is a solution of the motion equation of the fields which localized at the space-time. 
Time-dependent Lump ( as a dynamic Lump ) can radiate energy
and the static Lump cannot.
According to special relativity, an absence of the Lumps related to the massive or massless fields have different consequences.
Always, can be found a Lorentz transformation for the absence or existence of the massive Lump which can be generalized to the rest frame.
The glueballs and photoballs as the bound states get the mass, so we always find a Lorentz boost to reach to the rest frame.
Therefore, there is a one to one correspondence between the dynamic and static massive solutions. 
According to Coleman and Deser theorem, there is no finite - energy 
a non-singular solution of the classical Yang-Mills theory in four-dimensional Minkowski space-time which does not
radiates energy out to the spatial infinity.
In a linear classical field theory likes free electrodynamics, any non - singular initial a configuration of fields of finite energy will eventually spread out over all space.
The final state is outgoing radiation which independent of the initial configuration.
In contrast with dynamic Lump, there are non-linear field theories which radiate energy to spatial boundaries.
Also in Refs. \cite{coleman, coleman1}, the authors have shown that if in quantum field theory a static solution is forbidden then is forbidden in the classical limit.
That is the nonexistence of quantum glueballs say that nothing against the existence of classical glueballs \cite{coleman, coleman1, deser}.
\newline
The noncommutative physical theories are rewritten in term of noncommutative coordinates $[x^\m,x^\n]_\star=\i\theta^{\m\n}$ and as a nonlocal model does not hold the Lorentz symmetry \cite{Neto}.
The physics data forced us to study noncommutative version of physics further based on an antisymmetric tensor of the second rank: $\t$.
These studies have revealed some peculiar features of noncommutative quantum models.
In particular, much attention has also paid to
the noncommutative version of the Yang-Mills theory as well as noncommutative QED\cite{j1,j2,j4,j5,j6}.
The quantum electrodynamics as a massless theorem does not consist of the photon self - interaction.
Therefore, we do not encounter the photons bound states \cite{jafari}.
But, we can find similarity between nonabelian gauge theory and noncommutative electrodynamics.
Due to Refs.\cite{jafari, jafari1}, spatial and temporal noncommutative electrodynamics include a direct interaction term for photon-photon exchange. 
Therefore, massive bound state of quantum gauge bosons is mad by massive photons. 
So, the bound states of quantum fields in noncommutative electrodynamics,"\textit{photoballs}"
can exist.
In Ref.(\cite{jafari1}) and for the temporal noncommutativity $[x^0,x^1]=\i\t^{01}$, we have shown that for '$d=4$', $\hat{F}_{0i}=0$ that is the absence of electric fields is verified.
But, magnetic fields are free.
The purpose of the article is to research the existence of the solutions to Yang-Mills equation in 1+1 dimensional space-time.

\section{Presentation of the theory}
This work is written based on the Coleman and Deser approach.
We consider space-time to have one-time dimension and one space dimension.
Due to Ref.\cite{deser}, the vanishing of the energy-momentum tensor for the static systems excludes finite energy time-independent solutions of the sourceless theories.
Because the scalar current is trace-less ($\int\ d^{d-1}x\ T^{\m}_{\m}=0$) and this can be extracted directly from the equation of the motion.
This condition implies that the static solutions are forbidden.
To achieve the results of Ref.\cite{deser}, 
authors assumed that strength of the field must vanish at a far from the center of the Lumps.
Therefore, for the existence of the Lumps, the gauge field goes to zero in space boundaries because $\hat{F}^{\m\n}$ falls faster than $\mid \vec{r}\mid^{-\frac{1}{2}(d-1)}$.
This condition is necessary for the definition of localized fields.
But, we show here that can be questioned by a lower dimensional space-time. 
In contrast to Ref.\cite{deser}, we can say:

\textbf{Lemma}: \textit{In 1+1 dimensional space-time, the solution to Yang-Mills theory is possible.}

\textbf{proof}:
In four-dimensional Minkowski space-time, if $"g"$ stands for the coupling constant, then by introducing the localized and nonabelian strength tensor $F_a^{\m\n}=f_a^{\m\n}-\i g[A^\m,A^\n]_a$ with abelian form of the strength tensor of fields $f_a^{\m\n}=\p^\m A_a^\n-\p^\n A_a^\m$,
we have the beginning of the proof.
If the metric of two-dimensional space-time is similar to that of Minkowski $\eta_{\m\n}=diag(+1,-1,-1,-1)$, so
the Lagrangian which describes the theory of classical fields (gluons) becomes:
\begin{align}
\label{intial lagrange} 
\mathfrak{L}_{\mathrm{Nonabelian}}=-\frac{1}{2g^2}tr\{F_{\m\n}F^{\m\n}\}.
\end{align}
Here, all Greek indices run from 0 to 3 and the Latin indices indicate generators of the Lie algebra.
Also, $c^a_{bc}$ stands for the structure constants of the Lie algebra. 
In which, $[A_\m,A_\n]^a=\i c^a_{\ bc}A^b_\m A^c_\n$.
By variation of the action \cite{kleinert, parker}
the tensor of energy-momentum 
becomes:
\begin{align}
\label{energy - momentum}
T^{\m\n}_{\mathrm{Gluons}}=F^{\m\a}_{a}F^{a\ \b\n}\eta_{\a\b}-\frac{1}{4}\eta^{\m\n}F^{\a\b}_aF_{\a\b}^a.
\end{align}
That the factor of $\frac{1}{4}$ remains the same, regardless of the number of
dimensions is very important.
Definition $(\mathfrak{D}_\m)_a^{\ b}=\p_\m\delta^b_a+gc_a^{\ bc} A_{\m c}$, and by applying the least principle the equations of the motion of the fields are given by,
\begin{align}\label{meqof}
\p^\m F_{\m\n}^a+gc^a_{\ bc}A^b_\m F_{\m\n}^c=-gj^a_\n.
\end{align}
The tensor of energy-momentum obeys the following:
\begin{align}
\p_\m T^{\m\n}=0.
\end{align}
The $T^{\m\n}$ thus described is manifestly symmetric and traceless:
$$T^{\m\n} = T^{\n\m}$$ 
and 
$$T^\m_\m=0.$$

\textbf{The case of two-dimensional space-time}:
As the starting point, we have set the speed of light to unity.
We must say that the metric of two-dimensional space-time is:
\begin{align}\label{metric}
g_{\m\n}=\left(\begin{array}{cc}
1 &  0\\
0 &  -1
\end{array}
\right).
\end{align}
In Ref.(\cite{wheeler}), Wheeler shows that during the passage of the four-dimensional space-time into the two-dimensional, $p$ forms reduce to $p-1$ form.
Therefore, $\mathbf{A}$ as a $1$ form in four-dimensional space-time reduces to $0$ form $A$; $\ast\mathbf{A}\prec A$.
This rule will be extended to consists $2$ forms; $\ast F^{\m\n}\prec F^\a$.
Also, the current density $\mathbf{J}$  as $1$ form becomes a $0$ form; $\ast\mathbf{J}\prec j$.
Accordingly, two-dimensional space-time contains two electromagnetic subjects: $F_0=E$ and $F_1=B$ which based on the metric, these can take another form: $F^0=E$ and $F^1=-B$ \cite{wheeler}.
According to the language of differential forms, we can rewrite new tensor $G_\m$ in term of $F_\m$: $G_\m:=\e_{\m\n}F^\n$.
So that, $G_0=-B$ and $G_1=-E$.
Obviously, $F_0=\p_0 A$ and $F_1=\p_1 A$ satisfy Maxwell equations in two-dimensional space-time $\p_\m F^\m=j$ and $\p\wedge F=0$.
That is,
\begin{align}\label{moeqf2}
\p_1 F_1-\p_0 F_0=ej,
\cr
\p_0F_1-\p_1F_0=0.
\end{align}
In the case of Yang-Mills and for the case of $su(2)$ algebra,
the above statements take the form;
$F^a_0=E^a$ and $F^a_1=B^a$ relevant to 0 form potential $A^a$.
In which $a\in\{1,2,3\}$ and $[A,A]^a=\i\e^{abc}A_b A_c=0$.
So that $F^a_\m=f^a_\m$ without $ e\e^{abc}A_b A_c$. 
Also, Eq.(\ref{meqof}) reduces to Eq.(\ref{moeqf2}) in the new form:
\begin{align}\label{moeqf3}
\p_1 f^a_1-\p_0 f^a_0=ej^a,
\cr
\p_0f^a_1-\p_1f^a_0=0.
\end{align}
This show that each of the potential components takes an independent equation
without intertwining.  
The components of the strength tensor of fields $F^{\m\n}_a$ as $2$ forms defined on four-dimensional space-time 
will be a two-form $T^{\m\n}$ when space-time dimension reduces to two.
Due to Eq.(\ref{energy momentum}), the tensor of energy-momentum becomes:
\begin{align}
T_{\m\n}=\left(\begin{array}{cc}
F_0F_0 & F_0F_1\\
F_1F_0 & F_1F_1
\end{array}\right)-\frac{1}{2}(F_0F_0-F_1F_1)\left(\begin{array}{cc}
1 & 0 \\
0 & -1
\end{array}\right)
\end{align}
Therefore, when metric tensor is diagonal, the matrix elements of the energy-momentum tensor are:
\begin{align}
T=\left(
\begin{array}{cc}
\frac{1}{2}(E^2+B^2) & EB \\ 
BE & \frac{1}{2}(E^2+B^2)
\end{array}
\right).
\end{align}
Also, $T_{\m\n}=\frac{1}{2}(F_\m F_\n+G_\m G_\n)$, is a symmetric and trace less subject and satisfies:
$\p^\m T_{\m\n}=-jF_\n$.
Obtained statements get algebra's index for Yang-Mills case.
That is, $T_{\m\n}=\frac{1}{2}(F^a_\m F_{a\n}+G^a_\m G_{a\n})=\left(
\begin{array}{cc}
\frac{1}{2}(E^aE_a+B^aB_a) & E^aB_a \\ 
B^aE_a & \frac{1}{2}(E^aE_a+B^aB_a)
\end{array}
\right)=\Sigma_a T^a_{\m\n}$.

\textbf{Source free case:} $j=0$ denotes a state of the source-free case. 
So the equation of motion of stress tensor becomes:
$\p^\m T_{\m\n}=0$ which gives us three independent equations: $$\forall a\in su(2),\ \p^\m T^a_{\m\n}=0$$
We show that the vanishing of the energy-momentum tensor will be trivially and cannot create the additional condition on the fields.
Because, by dropping index of $"a"$, we can write,
\begin{align}
\p^\m T_{\m\n}=\p^\m(f_\m f_\n)-\frac{1}{2}\p^\m g_{\m\n}(f_\a f^\a)=0,
\end{align}
but from equation of motion of $f_\m;\ \p\cdot f=0$ and $\p\wedge f=0$, we have
\begin{align}
f_\m\p^\m f_\n-\frac{1}{2}\p_\n(f_\a f^\a)=0,
\end{align}
which for $\n=0$, last equation becomes:
\begin{align}
f_0\p^0 f_0-f_1\p_1f_0-\frac{1}{2}\p_0(f_\a f^\a)
\cr
\frac{1}{2}\p_0(f_0 f^0-f_1f_1)-\frac{1}{2}\p_0(f_\a f^\a)=0,
\end{align}
accordingly, the motion equation of energy-momentum tensor is trivial. 
Where in the last term we use from the tensor of metric.
Our process shows that only equation of motion of electrodynamics field in two-dimensional space-time is:
\begin{align}
\Box A_a=0,
\end{align}
where $\Box=\p^2_0-\p^2_1$.
Up to a constant function, $A_a$ has gauge freedom!
There is no lamp except with singularity, such as
\begin{align}\label{ssl}
A^a=A^a(x^1_0,x^0_0){\rm e}^{-k|x^1|-\omega x^0}.
\end{align}
With the condition: $k^2=\omega^2$ and singularity occurs at $x=0$.
Also, we can find a solution without singularity; $A^a=A^a(x^1_0,x^0_0){\rm e}^{-\i kx-\i\omega x^0}$.
The solution Eq.(\ref{ssl}) does not have an ability product the Lump without singularity.
So the motion equation of the energy-momentum tensor does not learn more than to the motion equation of the field. 
Consequently, for $1+1$ space-time dimensions, we learn nothing further from the energy-momentum tensor viewpoint.
Because of, they are not independent of each other.
The above statements confirm that there is a solution to QED in two-dimensional space-time but no Lump. 
Therefore the classical glueballs as a Lump of gluons cannot occur in two-dimensional space-time.
In fact, instead of the Ref.\cite{deser}, the solution to Yang-Mills theory is forbidden except in two and five-dimensional space-time.
Now, we must attend to $u_\star(1)$ on two-dimensional noncommutative space-time.
Noncommutativity on coordinates with the condition $[\hat{x}^0,\ \hat{x}^1]_\star=\imath\t^{01}$ is considered.
In this section, we suppose that all the employed functions are noncommutative coordinates dependent. 
Due to previous section, make substituting $g\rightarrow-\i e $, the $u_\star(1)$ action is
\begin{align} 
\label{nonc-action}
S_\star=-\frac{1}{4e^2}\int d^4 x\  (\hat{F}_{\star\m\n}\star \hat{F}_{\star\a\b})\eta^{\m\a}\eta^{\n\b},
\end{align}
where $\hat{F}^{\a\b}_\star=\p^{\a}A^{\b}_\star-\p^{\b} A^{\a}_\star-\imath e [A^{\a}_\star,\ A^{\b}_\star]_\star$ will be a function dependent on noncommutative coordinates, in which with 
$[A_\star,B_\star]_\star=A_\star\star B_\star-B_\star\star A_\star$, denotes the strength of the
noncommutative $u_{\star}(1)$ gauge fields in d-dimensional space-time.
They obey a different motion equation from $su(2)$; 
\begin{align}
\hat{D}_{\m}\star F^{\m\n}_\star=\p_{\m}F^{\m\n}_\star-\imath e[A_{\star\m},\ F^{\m\n}_\star]_{\star}=0,
\end{align}
where $\hat{D}_\m$ is a covarinat derivative includes $A_\star$.
Canonical energy-momentum tensor is given by
\begin{align}
\label{nonc-energy - momentum}
4T^{\m\n}_\star=2\{\hat{F}^{\m\a}_\star,\ \hat{F}^{\ \ \n}_{\star\a}\}_{\star}-\eta^{\m\n}(\hat{F}^{\a\b}_\star\star \hat{F}_{\star\b\a}),
\end{align}
here, $\{A_\star,B_\star\}_\star=A_\star\star B_\star+B_\star\star A_\star$.
In the usual form, the equation of motion of the energy-momentum tensor is 
\begin{align}
\label{nonc-motion equation}
\hat{D}_{\m}\star T^{\m\n}_\star=0.
\end{align}
\textbf{Two-dimensional noncommutative QED}
Due to the previous section, all $p$ forms defined on four-dimensional space-time reduce to $p-1$ forms when the dimension of space-time passage to two.
Also, $F^\m_\star\equiv f^\m_\star$ because $[A_\star,A_\star]_\star=0$.
In the case of source-free, this relevant equation of motion becomes;
\begin{align}
\p_1 f^1_\star+\p_0 f^0_\star-\i e[A_\star,f^1_\star]_\star=0,
\cr
\p_1 f^0_\star+\p_0 f^1_\star-\i e[A_\star,f^0_\star]_\star=0,
\end{align}
in which, in the brief form, 
\begin{align}\label{emof}
\hat{D}_\m\cdot f^\m_{\star}=0,
\cr
\hat{D}_\m\wedge f^\m_{\star}=0.
\end{align}
the above equations point to:
\begin{align}
-\Box A_\star-\i e[A_\star,\p_1 A_\star]_\star=0,
\cr
\p_0\p_1A_\star-\p_1\p_0A_\star-\i e[A_\star,\p_0A_\star]_\star=0.
\end{align}
So that we reach; $[A_\star,\p_0A_\star]_\star=0$.
This implies that $\p_1 f_{\star0}-\p_0 f_{\star1}=0$ and,
\begin{align}
\p_0A_\star=\left\{\begin{array}{c}
constant\\
\propto A_\star
\end{array}\right.
\end{align}
By choosing $\p_0A_\star=constant$, we will not get anything.
For the second choice, since we did not assume a source term for the electromagnetic fields so here is a perfect symmetry between space and time.
Accordingly, we can set $\p_1A_\star\propto A_\star$, and we find: 
\begin{align}
\Box A_\star=0.
\end{align}
Now, if the metric tensor is diagonal; Eq.(\ref{metric}), then the matrix elements of the energy-momentum tensor are:
\begin{align}
T^{\m\n}_\star=\frac{1}{2}\{f^\m_\star,f^\n_\star\}_\star-\frac{1}{2}g^{\m\n}(f^\n_\star\star f_{\star\n})
\cr
=\frac{1}{2}\{f^\m_\star,f^\n_\star\}_\star+\frac{1}{2}\{G^\m_\star,G^\n_\star\}_\star,
\end{align}
which obeys Eq.(\ref{nonc-motion equation}).
We can show directly that the motion equation of the energy-momentum tensor: $\p_0T^{0\n}_\star+\p_1T^{1\n}_\star-\i e[A_\star,T^{1\n}_\star]_\star=0$ can be obtained from the equation of motion of field Eq.(\ref{emof}).
Therefore, the $u_\star(1)$ is consistent with $\equiv su(2)$.
Consequently, there is a similar Lump with singularity, $A_\star=A(x^1_0,x^0_0){\rm e}^{-k|x^1|-\omega x^0}_\star$.
Namely, the bound state of noncommutative photons cannot occur in two-dimensional $u_\star(1)$. 

\section{Discussion}
In this paper, we study the existence of Lumps to Yang-Mills and $U_\star(1)$ (noncommutative quantum electrodynamics in the temporal mode) in 1+1 dimensional space-time.
We consider the gluon and photon self-interaction term which appear in Lagrangian density.
We derived the energy-momentum tensor and showed that we learn nothing further from the motion equation of the energy-momentum tensor.
Moreover, we show that if we use from the gauge freedom, then the solution to Yang-Mills and noncommutative quantum electrodynamics theories are possible.
Obtained results are instead to Deser and Coleman report.
Therefore, in 1+1 dimensional space-time, our calculations lead us to accept the following results: \\
1-"Glueballs cannot occur as Lump solution of the Yang-Mills theory".\\
2-"The existence of the photoballs is forbidden in $u_\star(1)$".\\
3-"The comprehensively similarity exists between $su(2)$ and $u_\star(1)$."
\section{Acknowledgments}
I would like to thanks Nicholas Wheeler, from his findings, appears in: "Electrodynamics in two-dimensional space-time, Reed College 1997". 
The author thanks the Shahrekord university for supporting this research grant fund.
\newline


\end{document}